\documentclass[journal=ancac3,manuscript=article]{achemso}

\usepackage{chemformula} 
\usepackage[T1]{fontenc} 

\author{Francesca Telesio}
\email{francesca.telesio@nano.cnr.it}
\affiliation{NEST, Istituto Nanoscienze-CNR and Scuola Normale Superiore, Piazza San Silvestro 12, 56127 Pisa, Italy}

\author{Elisa Passaglia}
\affiliation{Istituto di Chimica dei Composti Organometallici (CNR-ICCOM), SS Pisa, Via Moruzzi~1, 56124 Pisa, Italy}

\author{Francesca Cicogna}
\affiliation{Istituto di Chimica dei Composti Organometallici (CNR-ICCOM), SS Pisa, Via Moruzzi~1, 56124 Pisa, Italy}

\author{Federica Costantino}
\affiliation{Istituto di Chimica dei Composti Organometallici (CNR-ICCOM), SS Pisa, Via Moruzzi~1, 56124 Pisa, Italy}

\author{Manuel Serrano--Ruiz}
\affiliation{Istituto di Chimica dei Composti Organometallici (CNR-ICCOM), Via Madonna del Piano 10, 50019 Sesto Fiorentino, Italy}

\author{Maurizio Peruzzini}
\affiliation{Istituto di Chimica dei Composti Organometallici (CNR-ICCOM), Via Madonna del Piano 10, 50019 Sesto Fiorentino, Italy}

\author{Stefan Heun}
\email{stefan.heun@nano.cnr.it}
\affiliation{NEST, Istituto Nanoscienze-CNR and Scuola Normale Superiore, Piazza San Silvestro 12, 56127 Pisa, Italy}

\title{Hybrid 2D Black Phosphorus/Polymer Materials: New Platforms for Device Fabrication}


\begin{document}

\begin{tocentry}

\includegraphics{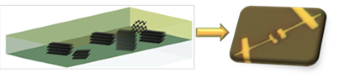}

Novel insight in hybrid materials is given. A judicious choice of the matrix polymer allows the exploitation of the full potential of the 2D filler: poly (methyl methacrylate) nanocomposites open a novel way towards a scalable, low cost production route for few-layer black phosphorus devices.

\end{tocentry}

\begin{abstract}
Hybrid materials, containing a 2D filler embedded in a polymeric matrix, are an interesting platform for several applications, because of the variety of properties that the filler can impart to the polymer matrix when dispersed at the nanoscale. Moreover, novel properties could arise from the interaction between the two. Mostly the bulk properties of these materials have been studied so far, especially focusing on how the filler changes the polymeric matrix properties. Here we propose a complete change of perspective by using the hybrid nanocomposite material as a platform suitable to engineer the properties of the filler and to exploit its potential in the fabrication of devices. As a proof of concept of the versatility and potentiality of the new method, we applied this approach to prepare black phosphorus nanocomposites through its dispersion in poly (methyl methacrylate). Black phosphorus is a very interesting 2D material, whose application have so far been limited by its very high reactivity to oxygen and water. In this respect, we show that electronic-grade black phosphorus flakes, already embedded in a protecting matrix since their exfoliation from the bulk material, are endowed with significant increased stability, and can be further processed into devices without degrading their properties.
\end{abstract}

\section{Introduction}

Hybrid materials, where a 2D filler ---an exfoliated van der Waals crystal--- is embedded in a polymeric matrix, have been widely studied in recent years, especially looking for an improvement of polymeric matrix properties because of the filler.\cite{Potts2011, Sahu, Kim2010a} Here, we propose a different approach to this class of hybrid materials, where a judicious choice of the right polymer in view of a specific application allows for a better exploitation of the full potential of the filler. 

Black phosphorus (bP) is a very suitable test bench for this approach. Since its ''re-discovery'' in 2014,\cite{Castellanos-Gomez2014b} many applications were proposed because of its interesting properties: band gap tunability with layer number (from 0.3~eV of the bulk crystal to about 2~eV for the monolayer),\cite{Das, Castellanos-Gomez2015, Morita1986} anisotropic transport properties in--plane,\cite{Xia2014a,Lee2015a} as well as high mobility, up to 45000~cm$^2$V$^{-1}$s$^{-1}$ for few-layer bP encapsulated in hexagonal boron nitride.\cite{Long2016} On the other hand, this material is very sensitive to oxygen, water, and light,\citep{Favron2015a, Ahmed2017, Huang2016, Luo2016, Castellanos-Gomez2014b} which represents a major drawback to its application in the few-layer form, especially since degradation issues become more and more relevant going towards thinner flakes.\cite{Favron2015a} 

Up to now, most of the results on few-layer bP have been obtained by preparing the devices under controlled atmosphere,\cite{Doganov2015, Long2016} which requires both technical efforts and sophisticated equipment, since a glove box is needed to properly preserve the material properties. Moreover, after exfoliation the few-layer bP needs an appropriate protection to undergo device fabrication and measurements. Several materials have been used so far for this purpose, ranging from other van der Waals materials \cite{Doganov2015, Doganov2015a} to inert oxides.\cite{Wood, Deng2017a, Galceran2017} Polymers have been as well identified as a low cost, easy, and efficient method to preserve multilayer bP quality,\cite{Jia2015a} and in particular poly (methyl methacrylate) (PMMA) has been widely used for this purpose \cite{Tayari2014}. This is very convenient, since PMMA is also the most common resist for electron beam lithography (EBL), which is a widely used technique to design prototypical devices for electrical, optical, and thermal transport applications. Since polymers efficiently prevent ambient degradation of bP, polymer nanocomposites could represent an efficient platform to stabilize the material against oxygen, moisture, and light irradiation effects. Up to now there are just few examples in literature of bP hybrid materials\cite{Passaglia2016, Sajedi-Moghaddam2017, Kumar2016, Li2016d}, where  bP is exfoliated by liquid phase exfoliation (LPE) and then mixed with a polymer, creating the hybrid material. In those cases, the polymer chains can interact with LPE few-layer bP preventing aggregation \cite{Passaglia2016}. This approach allowed multiple applications, ranging from biology,\cite{Kumar2016} electronics,\cite{Sajedi-Moghaddam2017} to non--linear optics,\cite{Li2016d} but requires a protective environment during the bP exfoliation step. Moreover the risk of contamination of the final composite with molecules of the high boiling solvents, typically used for bP LPE, cannot be excluded.\cite{Serrano-Ruiz2016,Sajedi-Moghaddam2017}

Here, we propose a completely different approach to the nanocomposite, focused on bP device fabrication, which allows the application of the most used fabrication techniques. bP nanosheets are produced by sonication in methyl methacrylate (MMA), as an efficient way to isolate electronic-grade bP flakes and the ultimate nanocomposite by in situ polymerization.\citep{Passaglia2018} Remarkably, this alternative fabrication approach does not require any protective environment, nor during exfoliation, neither during fabrication, which represents a significant simplification and improvement in the experimental procedure to prepare novel bP-based devices. Thus, simple devices may be easily produced from this material using standard EBL, with confirmed quality of the material itself, in terms of both Raman activity and  electrical transport properties. Our approach is a proof of concept of a very versatile platform for device engineering, with a large potential for the fabrication of different devices based on the use of sensitive 2D-materials, like bP, as it does not require any special protective environmental conditions for assuring bP stabilization.

\section{Results and discussion}

\subsection{Nanocomposite preparation and characterization}

The nanocomposite used in this work was prepared by sonication of a bP crystal, less reactive to oxygen and air moisture,\cite{Favron2015a, Huang2016} in MMA, followed by in situ polymerization, as previously described by us.\cite{Passaglia2018} In the case at hand, 0.0051 mg of crystalline bP were sonicated in 3.76~g of MMA for 90 min using a sonotrode. The hybrid material was obtained by in situ radical polymerization initiated by 2,2'-azobis(2-methylpropiolnitrile) (AIBN) 2 wt\% with respect to the monomer. Assuming that all bP participates in the composite formation, the bP concentration in the nanocomposite was approximately 0.2~\%.

The isolated composite was characterized by several techniques, including Raman spectroscopy, size exclusion chromatography (SEC) for evaluation of molecular weight, differential scanning calorimetry (DSC), and thermal gravimetric analysis (TGA) for determining thermal features.

The mass distribution, as well as the thermal properties of this nanocomposite were compared with a blank PMMA sample prepared under identical polymerization conditions. Both the number average molecular weight ($\overline{M\textsubscript{n}}=52000$~Da) and the weight average molecular weight ($\overline{M\textsubscript{w}}=178000$~Da) were higher compared to the blank sample (45000~Da and 103000~Da, respectively), indicating an effective role taken by the 2D black phosphorus during the polymerization process. In particular, the higher molecular weight suggests the growing of macromolecules in a ''confined space'' and thus near or between the bP layers.\cite{Passaglia2018} We measured, for the hybrid material, a glass transition temperature T$_g$ of 121.9$^\circ$C, while for the reference sample T$_g$ is 120.6$^\circ$C. From the TGA curve we extracted T$\textsubscript{onset}$ and T$\textsubscript{infl}$, determined as the temperature of intercept of tangents before and after the degradation step. For the nanocomposite we measured a T$\textsubscript{onset}$ of degradation of 270$^\circ$ C and T$\textsubscript{infl}$ of 293$^\circ$C - 372$^\circ$C (meaning the temperatures of maximum rate of degradation steps), while for the reference blank sample we measured   272$^\circ$C and 287$^\circ$C-381$^\circ$C, respectively. These measurements of thermal properties show a very similar behavior for the nanocomposite and the blank polymer sample. This, in turn, demonstrates a very good dispersion of the filler and suggests the formation of an interpenetrated phase, without any distinguishable effect on the bulk thermal properties.

\begin{figure}[t]
\centering
  \includegraphics[height=4.0cm]{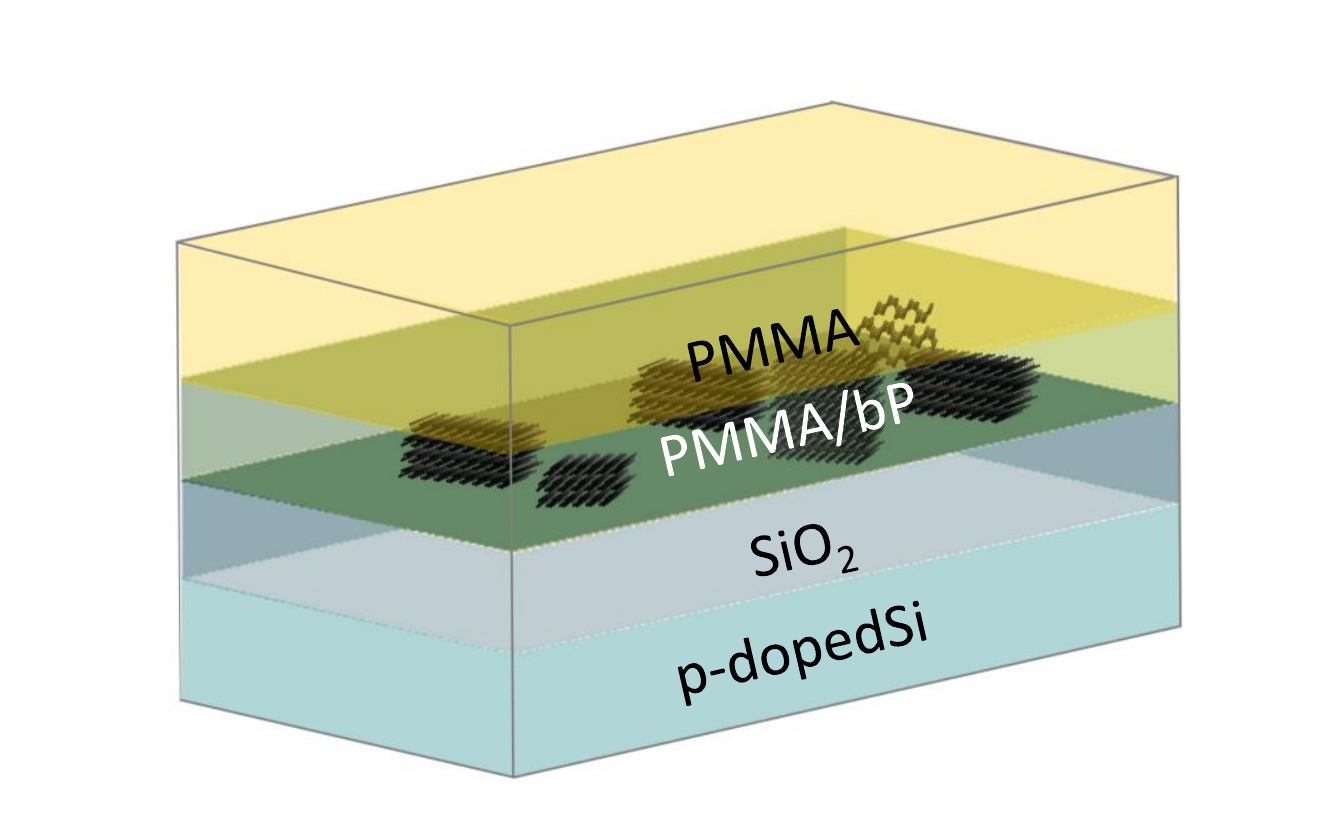}
  \caption{bP nanosheets embedded in PMMA, on a Si/SiO$_2$ substrate. The sketch is not to scale, so proportions are not respected. The SiO$_2$ layer is 300~nm thick, while the PMMA/bP layer is about 50~nm thick, and the PMMA layer 300~nm.}
  \label{fgr:fig1}
\end{figure}

\begin{figure*}[t]
\centering
  \includegraphics[height=7cm]{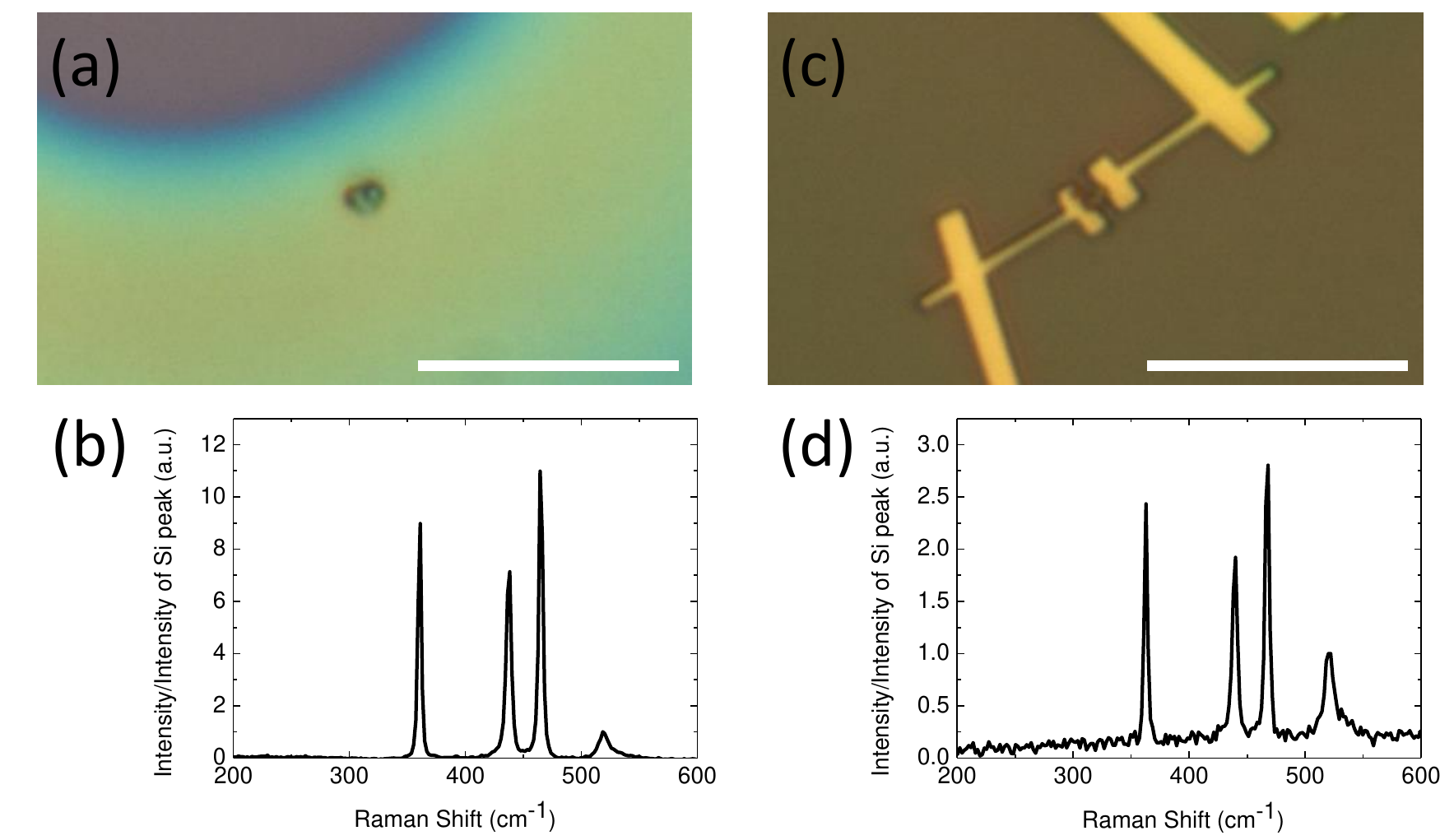}
  \caption{Optical microscopy and Raman spectroscopy performed on the device before [(a),(b)] and after [(c),(d)] fabrication. The scale bar in the optical microscopy images is 10 $\mu$m.}
  \label{fgr:fig2}
\end{figure*}

Then, 51.4 mg of the hybrid were dissolved under dry nitrogen in 2~mL of anisole and closed in a vial, before being spin-coated in ambient condition in a cleanroom environement on SiO$_2$/Si pre-patterned substrates, with Boron--doped Silicon and 300~nm of thermal SiO$_2$, as shown in Fig.~\ref{fgr:fig1}. In order to obtain a thin, homogeneous film, the spinner was set at 5000~revolutions per min, for 1 min. Then the samples were heated in air to $120^{\circ}$C to remove the solvent. On test samples, the resist thickness was measured with a stylus profilometer and was found to be approximately 50~nm. After this first process, the samples were analyzed by both optical microscopy and Raman spectroscopy to identify suitable nanosheets for fabrication. Nanosheet identification is facilitated by a marker grid previously patterned on the SiO$_2$/Si substrate by standard optical lithography. In Fig.~\ref{fgr:fig2} (a) and (b), an optical microscopy image of a small nanosheet, suitable for device fabrication, together with the corresponding Raman spectrum are shown. The three Raman peaks typical of bP, $A_g$, $B_{2g}$ and $A_g^2$,\cite{Sugai1985} are present. This is clear indication that small nanosheet structures such as the one shown in Fig. \ref{fgr:fig2}, with lateral size of a few $\mu$m, are still preserved, despite all processing made without any protective environmental conditions.
From different samples prepared by spin coating at different times, we could state that the bP/PMMA solution in anisole was stable for over 3 months and, on the same thin film sample, single individual nanosheets are still preserved as well for at least 3 months.

\subsection{Device fabrication}
The selected nanosheets were then processed by standard fabrication techniques. Thus, a 300~nm thick layer of bare commercial PMMA was spun on top of the thin layer of PMMA/bP nanocomposite, in order to have a sufficient PMMA thickness for a standard lithography and an easy lift-off process. Then electron beam lithography was performed. The possibility to achieve such a thin nanocomposite layer, if compared with the PMMA thickness commonly used for EBL, is a further important advantage of the proposed approach, since bP flakes are thus confined close to SiO$_2$ surface. After the lithographic process, the sample was developed and then underwent a very mild oxygen plasma treatment in vacuum, to remove resist residues and to allow for better ohmic contacts to the 2D material.\cite{Choi2011} Right after, the sample was placed in a thermal evaporator and pumped down. There, a bilayer composed of a stitching layer of nickel (10~nm) and an overlayer of gold (100~nm) was deposited. A standard lift-off process was performed in acetone (ACE) at 50$^{\circ}$C for 15 min. After rinsing with isopropanol (IPA) and drying under dry nitrogen flow, the sample was immediately coated by a bilayer of a methyl methacrylate methacrylic acid (MMA(8.5)MAA) copolymer and PMMA, in order to prevent the device from degrading during measurements. The final morphology of the device can be seen in Fig.~\ref{fgr:fig2} (c). In order to mount the sample onto the chip carrier for transport measurements, a further EBL step was performed, to open holes on the pads of the device, to allow wire-bonding from the device to the chip carrier. At the end of this multi-step fabrication process, the status of the bP nanosheet was checked by Raman (Fig.~\ref{fgr:fig2} (d)), which clearly showed all bP spectral features, demonstrating that bP quality was preserved during all fabrication steps. To evaluate  the quality of the material, the electrical transport properties of the device were then measured in vacuum both at room and at low temperature. 

\subsection{Electrical transport measurements}

\begin{figure*}[t]
\centering
  \includegraphics[height=6cm]{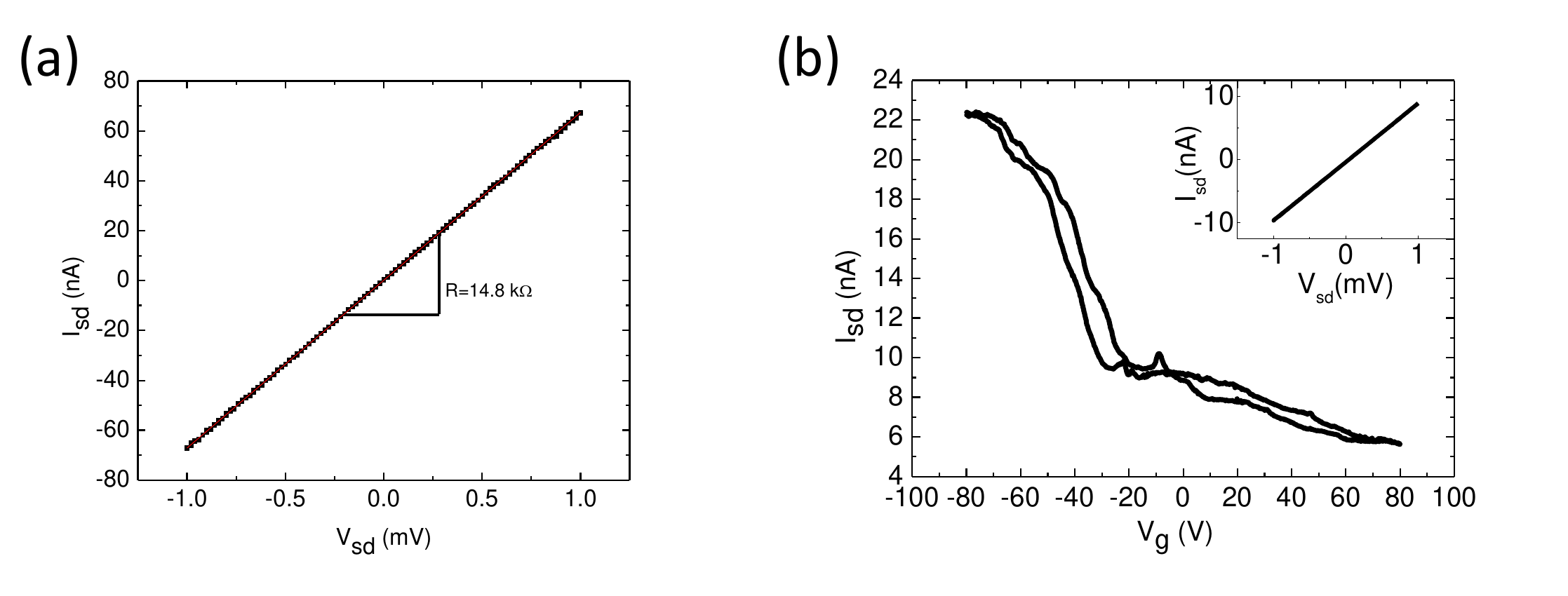}
  \caption{Transport properties of the 2-terminal device shown in Fig. \ref{fgr:fig2} (c), prepared from PMMA/bP nanocomposite. (a) Current versus voltage characteristics of the device at room temperature. (b) Low temperature measurements (4.2 K) Source drain current as a function of gate voltage for $V_{sd}$~=~1~mV. $I_{sd}-V_{sd}$ curve shown in the inset.}
  \label{fgr:fig3}
\end{figure*}

In Fig.~\ref{fgr:fig3} (a) the source--drain current-voltage ($I_{sd}-V_{sd}$) curve of the device at zero gate voltage (V$_g$) at room temperature is shown. The linear trend confirms an Ohmic behavior. The resistance obtained by fitting the $I_{sd}-V_{sd}$ curve is 14.8~k$\Omega$, compatible with exfoliated black phosphorus.\cite{Deng2017a} After a slow cool down in vacuum to 4.2~K, the resistance was measured again and was found to have increased by one order of magnitude,  to 108~k$\Omega$. By applying a back gate voltage, carrier concentration was modulated. The device shows a clear p-type semiconducting behavior, as expected for black phosphorus.\cite{Morita1986} We evaluate the field-effect mobility from the trans-conductance, as \cite{Du2014a, Gupta2006}
\begin{equation}
\mu_{FE} = \frac{dI_{sd}}{dV_g} \frac{L}{W} \frac{1}{C_{ox}V_{sd}}
\end{equation}
where $dI_{sd}/dV_g$ is the slope of the curve shown in Fig.~\ref{fgr:fig3}(b) extracted from the linear regime (between $-20$~V  and $-60$~V), $L$ is the device length, $W$ its width, $C_{ox}$ the capacitance of the 300~nm SiO$_2$ layer per unit of area, $C_{ox} = 11.5$~nF cm$^{-2}$. The extracted mobility value is 35~cm$^2$ V$^{-1}$s$^{-1}$ and this value, as well as sample resistance, is stable over 50 days. These results are compliant with the literature on electronic--grade liquid phase exfoliated bP flakes\citep{Kang2015} and comparable with mechanically exfoliated bP flakes on SiO$_2$.

\subsection{Further characterization}

\begin{figure}[t]
\centering
  \includegraphics[height=7cm]{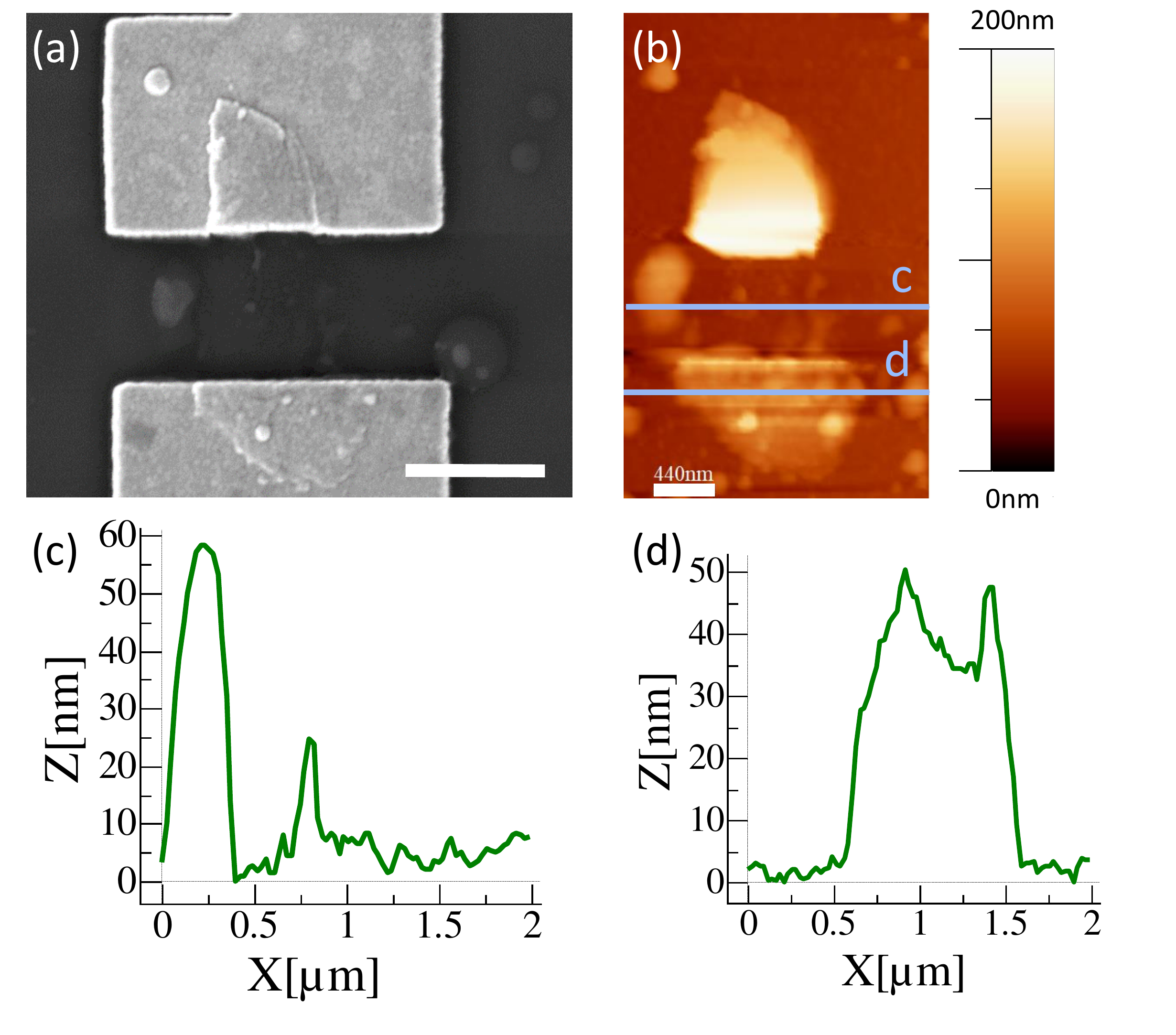}
  \caption{(a) SEM image of another device. The scale bar is 1~$\mu$m. (b) AFM of the same device, with two cross-sections (c) and (d). The scale bar of AFM image is 440~nm. The top layer of PMMA was removed in acetone and caused the accidental removal of the part of the device not clamped by the gold contacts, as shown in the central cross-section (c).}
  \label{fgr:fig4}
\end{figure}

As shown in Fig.~\ref{fgr:fig4} (a) and (b), SEM and AFM studies were carried out on a different device after the removal of the MMA(8.5)MAA copolymer/PMMA bilayer. \cite{nota1} Inspection of the AFM line profile shown in Fig.~\ref{fgr:fig4} (c) clearly shows that the central part of the flake, not clamped by gold, was accidentally removed during the cleaning with acetone. The upper and lower parts of the structure show height values between 50~nm and 100~nm (Fig.~\ref{fgr:fig4} (d)). This, together with morphological information collected on similar materials discussed in Passaglia et al.,\cite{Passaglia2018} suggests that the nanosheet is indeed an aggregate of smaller bP flakes, held together by thin polymer layers. On the other hand, our electrical transport measurements show that this material is comparable to electronic-grade liquid phase exfoliated bP material\cite{Kang2015} and suggest that bP in our hybrid material is of high quality. It can be expected that any improvement in preparation conditions will lead to even better transport properties, especially in terms of mobility.

\section{Conclusions}

In this work, we have proposed a new paradigm for hybrid 2D/polymer materials, focused on choosing the right polymer to exploit the properties of the 2D material at its best. We applied this approach to the preparation of a PMMA/bP nanocomposite. We showed that few--layer black phosphorus devices can be prepared from this multifunctional PMMA/bP hybrid material without the need for a glove box or any other kind of protective atmosphere, since the few-layer black phosphorus is exfoliated in MMA, which also efficiently acts as a protective coating. We showed also that few-layer bP survived the polymerization process of MMA without degrading. Furthermore, we demonstrated that the PMMA/bP nanocomposite is a suitable platform for device applications, since PMMA may be directly used as a resist for electron beam lithography. In fact, we used standard fabrication techniques to implement a simple device, and show a resistivity and carrier mobility characteristic of black phosphorus, as well as the expected p-type behaviour upon gate voltage modulation. The mobility can be further improved by optimizing sonication parameters. In summary, we have shown that the method here described constitutes a simple, low cost approach to handle sensitive 2D materials such as few-layer black phosphorus. Beyond that, the proposed approach can provide a series of innovative platforms by varying the 2D material and/or the embedding polymer, which can be tuned to provide other functionalities to the final composite.

\section{Methods}

\subsection{Experimental techniques}

Number average molecular weight ($\overline{M\textsubscript{n}}$) and weight average molecular weight ($\overline{M\textsubscript{w}}$) were determined using Size Exclusion Chromatography (SEC), Agilent Technologies 1200 Series. The instrument is equipped with an Agilent degasser, an isocratic HPLC pump, an Agilent refractive index (RI) detector, and two PLgel 5 $\mu$m MiniMIX-D columns conditioned at 35$^\circ$C. Chloroform (CHCl$_3$) was used as the mobile phase at a flow rate of 0.3~mL~min$^{-1}$. This system was calibrated with polystyrene standards in a range from 500 to $3 \times 10^5$~g~mol$^{-1}$. Samples were dissolved in CHCl$_3$ (2~mg~mL$^{-1}$) and filtered through a 0.20 micron syringe filter before analysis (twice in the case of hybrids). Number average molecular weight ($\overline{M\textsubscript{n}}$) and weight average molecular weight ($\overline{M\textsubscript{w}}$) were calculated using the Agilent ChemStation software.
 
Thermal Gravimetric Analyses (TGA) were carried out with a Seiko EXSTAR 7200 TGA/DTA by introducing about 5-8 mg of sample in an alumina sample pan of 70 $\mu$L. In a typical experiment, run was carried out at a standard rate of 10$^\circ$C/min from 30$^\circ$C to 700$^\circ$C under nitrogen flow. T$\textsubscript{onset}$ and T$\textsubscript{infl}$, which are the  were determined by analyzing the TGA curve (as the temperature of intercept of tangents before and after the degradation step) and DTG curve (as the maximum of the peak), respectively.
 
The glass transition temperature (T$\textsubscript{g}$) was determined by Differential Scanning Calorimetry (DSC) using a PerkinElmer DSC4000 equipped with intracooler and interfaced with Pyris software (version 9.0.2). The range of temperatures investigated was 40-180 $^\circ$C. Thermal scans were carried out on 5-10 mg samples in aluminum pans under nitrogen atmosphere. The instrument was calibrated by the standards In (Tm=156.6 $^\circ$C, $\Delta$ Hm=28.5 J/g) and Pb (Tm= 327.5$^\circ$C, $\Delta$Hm=23.03 J/g).

Raman spectroscopy was performed using a Renishaw inVia system equipped with a 532~nm laser and a motorized stage for 2D mapping of samples. A laser spot size of approximately 1~$\mu$m in diameter was used. Laser power was 55~$\mu$W. We have verified that this power did not damage the bP flakes during measurements. We note that PMMA is a suitable polymer for this kind of analysis, since it is transparent in the spectral region of interest.

For device fabrication, the commercial PMMA AR-P 679.04 was used as a positive  resist for electron beam lithography. Electron Beam Lithography (EBL) was performed with a Zeiss UltraPlus Scanning Electron Microscope (SEM), equipped with an interferometric stage for better alignment, and with a Raith tool for EBL. An acceleration voltage of 20~kV and a dose of 350~$\mu$C/cm$^2$ were used to expose the active area of the device. For the protective layer after fabrication, the commercial MMA(8.5)MAA copolymer (EL13) and PMMA (A4), both from MicroChem, were used. The development occurred in AR 600-56, a commercial developer from Allresist GmbH, based on 4-Methylpentan-2-one diluted in IPA.

Oxygen plasma treatment is performed at 10 W for 1 min, with a flux of 40 standard cubic centimeters per min (SCCM). 

Metal evaporation was performed in a Sistec multi-crucible thermal evaporator equipped with a rotating carousel, in order to have the sample aligned with the crucible of the selected metal. Evaporation rate was  0.4 \AA s$^{-1}$ for nickel and 1.5 \AA s$^{-1}$ for gold. The pressure in the evaporator chamber was around $10^{-6}$~mbar before evaporation, and  increased slightly during metal evaporation. 

For transport measurements, the samples were bonded using Au wire to a 16 pin dual in line chip carrier. The transport properties were measured in DC in vacuum ($p < 10^{-4}$~mbar) in a custom made insert, equipped with a diode for temperature measurement, and compatible with the used chip carriers. The leakage current though the gate was measured during $V_g$ loops and found to be negligible.

SEM imaging was performed with a Zeiss Merlin microscope with 5~kV acceleration voltage. Atomic force microscopy (AFM) measurements were performed with a Bruker Dimension Icon AFM, in pick force mode. Data  analysis was performed with WSxM software.\cite{Horcas2007}

\subsection{Materials}

In our experiments, we used bP crystals prepared according to the procedure developed by Nilges et al.,\cite{Nilges2008} wherein high-purity red phosphorus (>~99.99\%), tin (>~99.999\%), and gold (>~99.99\%) are heated in a muffle oven with a SnI$_4$ catalyst. The solid product was placed in a quartz tube, subjected to several evacuation-purge cycles with N$_2$ gas, and then sealed under vacuum. The evacuated quartz tube was heated to $406^{\circ}$C at $4.2^{\circ}$C/min, where it remained for 2 hours. The tube was then heated to $650^{\circ}$C at $2.2^{\circ}$C/min and held at this temperature for 3 days. The tube was then cooled to room temperature at $0.1^{\circ}$C/min. The final product is crystalline bP with a typical size of some mm. All the other materials (polymers, reagents, solvents) are commercial and used as they are without further purification.

\begin{acknowledgement}
This work was financially supported by EC through the project PHOSFUN {\it Phosphorene functionalization: a new platform for advanced multifunctional materials} (ERC ADVANCED GRANT No. 670173 to M.~P.). F.~T. thanks CNR-Nano for funding the SEED project SURPHOS. E.~P. thanks CNR for funding the project Ma.Po.Fun. (DCM.AD002.239). S.~H. thanks Scuola Normale Superiore for support, project SNS16\_B\_HEUN -- 004155. Funding from the European Union's Horizon 2020 research and innovation programme under grant agreement No. 696656 -- GrapheneCore1 is acknowledged. Financial support from CNR in the framework of the agreements on scientific collaborations between CNR and CNRS (France), NRF (Korea), and RFBR (Russia) is also acknowledged. We further acknowledge funding from the Italian Ministry of Foreign Affairs, Direzione Generale per la Promozione del Sistema Paese (agreement on scientific collaboration between Italy and Poland). Serena Coiai is acknowledged for helpful discussion. 
\end{acknowledgement}

\bibliography{PMMA_bP_2018-01-19}

\begin{mcitethebibliography}{36}
\bibitem[not()]{nota1}
MMA(8.5)MAA copolymer/PMMA bilayer would be exposed in case of SEM imaging and
  would not allow the height determination by AFM\relax
\mciteBstWouldAddEndPuncttrue
\mciteSetBstMidEndSepPunct{\mcitedefaultmidpunct}
{\mcitedefaultendpunct}{\mcitedefaultseppunct}\relax
\EndOfBibitem
\end{mcitethebibliography}

\end{document}